%% file: paper.tex
\documentclass[]{elsart}
\usepackage{amsmath}
\usepackage{amssymb}
\usepackage{graphicx}

\def\issue(#1,#2,#3){#1 (#3) #2} 

\def\opcit(#1){ {\em op. cit.}, #1}

\def\ARNPS(#1,#2,#3){Ann.\ Rev.\ Nucl.\ Part.\ Sci.\ \issue(#1,#2,#3)}
\def\CPC(#1,#2,#3){Comp.\ Phys.\ Comm.\ \issue(#1,#2,#3)}
\def\CIP(#1,#2,#3){Comput.\ Phys.\ \issue(#1,#2,#3)}
\def\EPJC(#1,#2,#3){Eur.\ Phys.\ J.\ C\ \issue(#1,#2,#3)}
\def\IEEETNS(#1,#2,#3){IEEE Trans.\ Nucl.\ Sci.\ \issue(#1,#2,#3)}
\def\NP(#1,#2,#3){Nucl.\ Phys.\ \issue(#1,#2,#3)}
\def\NIM(#1,#2,#3){ Nucl.\ Instrum.\ and Meth.\ \issue(#1,#2,#3)}
\def\PL(#1,#2,#3){Phys.\ Lett.\ \issue(#1,#2,#3)}
\def\PRD(#1,#2,#3){Phys.\ Rev.\ D \issue(#1,#2,#3)}
\def\PRL(#1,#2,#3){Phys.\ Rev.\ Lett.\ \issue(#1,#2,#3)}
\def\SJNP(#1,#2,#3){Sov.\ J. Nucl.\ Phys.\ \issue(#1,#2,#3)}
\def\ZPC(#1,#2,#3){Z.\ Phys.\ C \issue(#1,#2,#3)}

\begin{document}

\begin{frontmatter}

\title{Measurement of the doubly Cabibbo suppressed decay $D^0 \rightarrow K^+\pi^-$ and a 
search for charm mixing}

\input{authors}



\begin{abstract}
We present an analysis of the decay $D^0\!\rightarrow\!K^+\pi^-$ based on FOCUS data.
From a sample of $234$ signal
events, we find a branching ratio of 
$\frac{\Gamma(D^0\!\rightarrow\!K^+\pi^-)}{\Gamma(D^0\!\rightarrow\!K^-\pi^+)} = 
(0.429^{\,+\,0.063}_{\,-\,0.061}\pm 0.027)\%$ under the assumptions of no mixing and no \textit{CP} 
violation.  Allowing for \textit{CP} violation, we find a branching ratio of 
$(0.429 \pm 0.063 \pm 0.028)\%$ and a \textit{CP} asymmetry of 
$0.18 \pm 0.14 \pm 0.04$.  The branching ratio for the case of mixing with no 
\textit{CP} violation is $(0.381^{\,+\,0.167}_{\,-\,0.163}\pm0.092)\%$.  We also present limits
on charm mixing.
\end{abstract}


\end{frontmatter}

\section{Introduction}

While mixing in the strange and beauty quark sectors has been observed for many years, charm mixing
remains elusive.  In the Standard Model the charm mixing rate is greatly suppressed by small CKM matrix
elements and strong GIM suppression.  It is this very fact, however, that provides a unique opportunity
to search for new physics.

This paper represents a continuation of mixing studies by FOCUS begun with a measurement of $y_{CP}$ looking
for a lifetime difference between \textit{CP} even and mixed \textit{CP} states~\cite{focus_ycp} and
continued with an analysis of the decay $D^0\!\rightarrow\!K^+\pi^-$~\cite{linkpaper}.  A summary of charm 
mixing can be found in the recent Particle Data Group review~\cite{pdg}.

The search for charm mixing requires tagging the flavor of a neutral charm meson at production and 
tagging the flavor again at decay.  The production flavor is determined using $D^{*+}\!\rightarrow\!D^0\pi_s^+$
decays.  The charge of the slow pion $(\pi_s)$ determines the flavor of the produced neutral $D$ meson.  The flavor of the
decaying $D^0$ is determined by reconstructing a Cabibbo favored (CF) decay with a charged kaon such as 
$D^0\!\rightarrow\!K^-\pi^+$.  A right-sign (RS) decay is defined as one in which the kaon and soft pion charges are 
opposite while same kaon and $\pi_s$ charges are wrong-sign (WS) decays.  The $D^0$ can also decay directly to 
$K^+\pi^-$ via a doubly Cabibbo suppressed decay (DCSD).  In the small mixing limit (known to be true experimentally),
the time dependent wrong-sign decay rate can be written as:
\begin{equation}
\label{eq:wsrate}
R_\textrm{WS}(t) = e^{-\Gamma t}\left( R_D + \sqrt{R_D} y' \Gamma t + \frac{1}{4}\left({x'}^2+{y'}^2\right) \Gamma^2 t^2\right)
\end{equation}
where the three terms correspond to DCSD, interference between mixing and DCSD, and mixing.
$R_D$ is the DCS branching ratio relative to the Cabibbo favored mode.  The mixing parameters
$x' \equiv x \cos{\delta_{K\pi}} + y \sin{\delta_{K\pi}}$ and $y' \equiv y \cos{\delta_{K\pi}} - x \sin{\delta_{K\pi}}$ 
are rotated versions of the mixing parameters $x \equiv \frac{\Delta M}{\Gamma}$ and 
$y \equiv \frac{\Delta \Gamma}{2\Gamma}$ which are the mass and lifetime
splitting terms, respectively.  The angle $\delta_{K\pi}$ is the strong phase between the CF and DCS decay, 
generally expected to be small.  Analogously to the DCSD rate, $R_M \equiv \left({x'}^2+{y'}^2\right)/2$ is 
the time independent mixing rate. Discovery of hadronic charm mixing requires separating the three components 
of Eq.~\ref{eq:wsrate} using the different lifetime distributions.  Note also that Eq.~\ref{eq:wsrate} only 
has a ${x'}^2$ and therefore the sign of $x'$ cannot be determined from this analysis and the relevant fit 
variable is ${x'}^2$.

\textit{CP} violation can in principle occur in any of the three wrong-sign components resulting in three
additional terms, $A_D$, $A_M$, and $\phi$.  $A_D$ and $A_M$ are \textit{CP} asymmetries associated with 
the DCSD and mixing terms, respectively.  The \textit{CP} violation term associated with the interference 
simply rotates $\delta_{K\pi}$ by $\pm\phi$.  One choice for the \textit{CP} violating time dependent 
wrong-sign rate is:
\begin{equation}
\label{eq:wsratecp}
\begin{split}
R^{\pm}_\textrm{WS}(t) = &  \;e^{-\Gamma t} \biggl( R_D\sqrt{\frac{1 \pm A_D}{1 \mp A_D}} \\
&+\; \sqrt{R_D}\, \sqrt[4]{\frac{\left(1 \pm A_D\right)\left(1 \pm A_M\right)}{\left(1 \mp A_D\right)\left(1 \mp A_M\right)}}
\left(y'\!\cos{\phi}\mp x'\!\sin{\phi}\right)\Gamma t \\
&+\; \frac{1}{4} \sqrt{\frac{1 \pm A_M}{1 \mp A_M}} \!\left({x'}^2 + {y'}^2\right)\!  \Gamma^2 t^2 \biggr) \\
\end{split}
\end{equation}
where in all cases the upper (lower) sign refers to an initial $D^0$ $\left(\overline{D}\,\!^0\right)$.
We can use Eq.~\ref{eq:wsratecp} on the full $D^0 + \overline{D}\,\!^0$ data set to directly 
extract the six parameters of interest, $R_D$, ${x'}^2$, $y'$, $A_D$, $A_M$, and $\phi$.  There is a four-fold
degeneracy in Eq.~\ref{eq:wsratecp}. A 2-fold degeneracy appears when
$x'$, $y'$, and $\phi$ are replaced with $-x'$, $-y'$, and $\phi +
\pi$ which does not change Eq.~\ref{eq:wsratecp}.  Requiring
$|\phi|<\pi/2$ removes this degeneracy and follows the convention of BABAR~\cite{babar_hmix}.  
Operationally, this is done by replacing 
$\cos{\phi}$ with $\sqrt{1 - \sin^2{\phi}}$ and fitting for
$\sin{\phi}$ rather than $\phi$.  The other 2-fold degeneracy occurs when 
replacing $x'$ and $\phi$ with $-x'$ and $-\phi$, again leaving
Eq.~\ref{eq:wsratecp} unchanged. We remove this degeneracy
by replacing $x'$ with $|x'|$.  This convention avoids a specious dependence on the sign of $x'$.

An alternative method for extracting the six parameters of interest is to use Eq.~\ref{eq:wsrate} on 
the $D^0$ and $\overline{D}\,\!^0$ samples separately and obtain values of $R_D$, ${x'}^2$, and $y'$ for 
each.  We obtain $\left\{R_D^+, {x'}^{+^2}, {y'}^{+}\right\}$ and 
$\left\{R_D^-, {x'}^{-^2}, {y'}^{-}\right\}$ from the $D^0$ and $\overline{D}\,\!^0$ samples, respectively.  
With $R_M^\pm \equiv  \left({x'}^{\pm^2}+{y'}^{\pm^2}\right)/2$, the \textit{CP} violation asymmetries 
can be found as:
\begin{equation}
\label{eq:acp}
A_D \;=\; \frac{R_D^+ - R_D^-}{R_D^+ + R_D^-} \hspace{20pt} \textrm{and} \hspace{20pt} A_M \;=\; 
\frac{R_M^+ - R_M^-}{R_M^+ + R_M^-}.
\end{equation}
while ${x'}^\pm$ and ${y'}^\pm$ are related by:
\begin{eqnarray}
\label{eq:xrot}
{x'}^\pm & \;=\; & \sqrt[4]{\frac{1 \pm A_M}{1 \mp A_M}} \left({ x'}\!\cos{\phi}\pm{ y'}\!\sin{\phi}\right) \\
{y'}^\pm & \;=\; & \sqrt[4]{\frac{1 \pm A_M}{1 \mp A_M}} \left({ y'}\!\cos{\phi}\mp{ x'}\!\sin{\phi}\right).
\label{eq:yrot}
\end{eqnarray}
Choosing either the $+$ or $-$ equations from Eqs.~\ref{eq:xrot},\ref{eq:yrot}, we can square both sides 
and add the equations together.  The result, along with rewriting Eq.~\ref{eq:yrot}, gives three relations:
\begin{eqnarray}
\label{eq:ypslv1_babar}
{y'}\cos{\phi} - {x'}\sin{\phi} & \;=\; & {y'}^+\sqrt[4]{\frac{1-A_M}{1+A_M}} \;=\; {y'}^+\sqrt[4]{\frac{R_M^-}{R_M^+}} \\
{y'}\cos{\phi} + {x'}\sin{\phi} & \;=\; & {y'}^-\sqrt[4]{\frac{1+A_M}{1-A_M}} \;=\; {y'}^+\sqrt[4]{\frac{R_M^+}{R_M^-}} \\
\frac{{x'}^2 + {y'}^2}{2} & \;=\; & R_M^+\sqrt{\frac{1-A_M}{1+A_M}} =\; R_M^-\sqrt{\frac{1+A_M}{1-A_M}} \,=\; \sqrt{R_M^+ R_M^-}.
\label{eq:xpypslv_babar}
\end{eqnarray}
Again replacing $\cos{\phi}$ with $\sqrt{1-\sin^2{\phi}}$, one can solve these equations for 
$x'^2$, $y'$, and $\sin{\phi}$.  There are two solutions corresponding to the relative sign between 
${x'}^+$ and ${x'}^-$.  
In this framework it is natural to define $R_D \equiv \sqrt{R_D^+R_D^-}$.

\section{Event reconstruction and selection}

The FOCUS experiment recorded data during the 1996--7 fixed-target run at Fermilab.  
A photon beam obtained from bremsstrahlung of 300\,GeV electrons and positrons impinged
on a set of BeO targets.  Four sets of silicon strip detectors, each with three views, 
were located downstream of the targets for vertexing and track finding.  For most of 
the run, two pairs of silicon strips were also interleaved with the target segments for more
precise vertexing~\cite{tsilicon}.  Charged particles were tracked and momentum analyzed as they passed
through up to two dipole magnets and up to five sets of multiwire proportional chambers 
with four views each.  Three multicell threshold \v{C}erenkov counters, two electromagnetic
calorimeters, and two muon detectors provided particle identification.  A hadronic trigger 
passed 6 billion events for reconstruction.  The average photon energy of reconstructed charm
events is 175\,GeV and the average $D^0$ momentum for this analysis is 75\,GeV/$c$.

A candidate driven vertexing algorithm is used to reconstruct charm.  In the case of
$D^0\!\rightarrow\!K^-\pi^+$, two oppositely charged tracks are required to verticize with
CL $>$ 2\%.  The momentum and vertex location are used as a ``seed'' track to find the 
production vertex which must have CL $>$ 1\%.  The flavor of the produced $D$ meson
is determined using the decay $D^{*+} \rightarrow D^0\pi_s^+$.  The soft pion must
be consistent with originating from the production vertex and the track is refit using the production
vertex as an extra constraint.  The energy release 
($Q(D^*) \equiv M(D^*)-M(D^0)-m_\pi$) must be less than $55\,\textrm{MeV}/c^2$.
Separating charm from hadronic background is primarily accomplished by requiring the decay
vertex be distinct from the production vertex.  A loose cut of $\ell/\sigma_\ell > 2$ is applied
where $\ell$ is the distance between the two vertices and $\sigma_\ell$ is 
the calculated uncertainty ($<$$\sigma_\ell$$>$ $\sim$ 500 $\mu$m).
Since hadronic reinteractions can fake a decay, requiring the secondary vertex to be located
outside of target material reduces background.  The out-of-material significance $\sigma_\textrm{out}$ is
positive (negative) for a vertex outside (inside) material.  We require 
$2\,\ell/\sigma_\ell + \max{\left(-2,\sigma_\textrm{out}\right)} > 6$.  To ensure the $D^0$ decay tracks
do not originate from the production vertex, a cut is made on the change in production vertex confidence
level when either $D^0$ decay track is added to the vertex ($\Delta \textrm{CL}_\textrm{pri}$).  
We require $\Delta \textrm{CL}_\textrm{pri} < 20\%$.  Since the signal contains a charged kaon,
information from the three \v{C}erenkov counters effectively suppresses backgrounds.  The 
\v{C}erenkov identification algorithm~\cite{citadl} returns negative log-likelihood values 
$\mathcal{W}_i(j)$ for track $j$ and hypothesis $i\in\left\{e,\pi,K,p\right\}$.  In practice,
differences in log-likelihoods between hypotheses are used such as 
$\Delta\mathcal{W}_{\pi K} \equiv \mathcal{W}_\pi-\mathcal{W}_K$.  We require 
$\Delta\mathcal{W}_{\pi K}(K)>0.5$, $\Delta\mathcal{W}_{K\pi}(\pi)>-3$, 
$\mathcal{W}_\textrm{min}(\pi)-\mathcal{W}_\pi(\pi) > -5$, and
$\mathcal{W}_\textrm{min}(\pi_s)-\mathcal{W}_\pi(\pi_s) > -5$ where 
$\mathcal{W}_\textrm{min} \equiv \min{(\mathcal{W}_{i\in\{e,\pi,K,p\}})}$.  We also require
$\Delta\mathcal{W}_{\pi K}(K) + \Delta\mathcal{W}_{K\pi}(\pi) > 3$.  To suppress doubly misidentified
$D^0\!\rightarrow\!K^-\pi^+$ decays (reconstructed as $D^0\!\rightarrow\!\pi^-K^+$), we require 
$\Delta\mathcal{W}_{\pi K}(K) + \Delta\mathcal{W}_{K\pi}(\pi) > 8.5 - 0.5 \left[M_N^\textrm{ref}(D^0)\right]^2$ 
where $M_N^\textrm{ref}(D^0)$ is the normalized reflected mass obtained by switching the rest mass values
for the $\pi$ and $K$, subtracting the nominal $D^0$ mass, and dividing by the calculated error.  
The remaining selection criteria are very efficient and mildly suppress some
backgrounds.  To suppress semileptonic decays we require the $\pi$ candidate not be
identified as a muon by the muon detectors.  Both pions must be identified as not being an electron by either the 
\v{C}erenkov system $\left(\Delta\mathcal{W}_{e\pi}(\pi)>1\right)$ or a calorimeter.  All tracks used in the
analysis
must be inconsistent with the beam direction to suppress tracks from $\gamma \!\rightarrow\! e^+e^-$.
Momentum requirements of $p(K),p(\pi) > 6\,\textrm{GeV}/c$ and $p(\pi_s) > 2\,\textrm{GeV}/c$ are set.
The calculated uncertainties on the lifetime and mass must be less than $150\,\textrm{fs}$ and 
$25\,\textrm{MeV}/c^2$, respectively.
The momentum asymmetry, $\left|\frac{p(K)-p(\pi)}{p(K)+p(\pi)}\right|$, must be
less than $\frac{p(D)+100\,\textrm{GeV}/c}{200\,\textrm{GeV}/c}$.  The summed $p_T^2$
of $D^0$ daughters with respect to the $D^0$ momentum vector must be greater than $0.25\,\textrm{GeV}^2/c^2$.
These last two cuts favor a decay of a heavy particle over combinatoric background.
If, after passing all these cuts, more than one $D^*$ candidate is found for one $D^0$ candidate, an additional
cut may be made.
If a right-sign (wrong-sign) candidate is found with $2.5\,\textrm{MeV}/c^2 < Q(D^*) < 9.5\,\textrm{MeV}/c^2$, 
then no wrong-sign (right-sign) candidates are allowed.

\section{Mixing fit}

A three dimensional binned maximum likelihood fit is used in this analysis.  Two of the dimensions, 
$M(D^0)$ and $Q(D^*)$, are useful for separating signal from background while the third dimension,
$\tau(D^0)$, is necessary to distinguish DCSD, mixing, and interference contributions.  The $D^0$ mass
is divided in 30 equal bins from 1.75 to 2.05 GeV/$c^2$.  The energy release is divided into 110 equal bins
from 0 to 55 MeV/$c^2$.  The $D^0$ proper lifetime is divided into 5 bins of 0.20\,ps from 0 to 1\,ps,
one bin from 1.0--1.25\,ps, one bin from 1.25--1.75\,ps, and one bin from 1.75--3\,ps.  These bins are 
much larger than our proper time resolution of $\sim$35\,fs.  The right-sign
and wrong-sign data are fit simultaneously.  The negative log-likelihood $\ell \equiv -\log{\mathcal{L}}$
is simply the sum of the negative log-likelihoods of each bin, $\ell = \sum_i^{N_\textrm{bins}}{\ell_i}$.  
\textsc{Minuit}~\cite{minuit} is used to minimize $\ell$ and $1$-$\sigma$ errors are defined as the point
where $\ell$ changes by $0.5$ with respect to the minimum value using the \textsc{Minos}~\cite{minuit} 
algroithm.

The fit has many components.  These components generally have a 3-D shape determined by a 
Monte Carlo simulation and a yield which is free to float, some with weak constraints imposed.  The 
components making up the fit which are modeled by the Monte Carlo are RS signal, 
WS signal (DCSD, interference, and mixing), real 
$D^0\!\rightarrow\!K^-\pi^+$ decay with a fake soft pion,
reflections to both RS and WS ($K^-K^+$, $\pi^-\pi^+$, $\pi^-\pi^+\pi^0$, $K^0\pi^-\pi^+$), RS background
($K^-\ell^+\nu$ and $K^-\pi^+\pi^0$), and WS background (double misidentification of $K^-\ell^+\nu$, 
$K^-\pi^+\pi^0$, $K^-\pi^+$).  The 3-D ($M(D^0)$, $Q(D^*)$, \&
$(\tau(D^0)$) shapes are obtained from a Monte Carlo simulation with
the PDG~\cite{pdg} value for the $D^0$ lifetime (appropriately
modified for WS signal mixing and interference terms).
The remaining background, labeled \texttt{random}, is some combination of non-charm and poorly
reconstructed charm
events.  In the default fit, this contribution is modeled with functional forms in mass 
$a \exp{\left(b m\right)}$, energy release $\alpha q^{1/2} + \beta q^{3/2}$, and lifetime 
$\exp{\left(-t/\tau_1\right)} + \eta \exp{\left(-t/\tau_2\right)}$
where $a$, $b$, $\alpha$, $\beta$, $\eta$, $\tau_1$, and $\tau_2$ are
free parameters of the fit.  In one fit variation, this 
contribution is modeled instead as a combination of minimum bias and generic charm events (without the 
decay modes accounted for above) from Monte Carlo.

Penalty terms are added to the likelihood to ensure that many of the backgrounds described above
are consistent with known branching ratios.  The affected background components are the $D^0$ decays
to $K^+K^-$, $\pi^+\pi^-$, $\pi^+\pi^-\pi^0$, $K^0\pi^+\pi^-$, $K^-\pi^+\pi^0$, and $K^-\ell^+\nu$.  These
six reflections appear in both right-sign and wrong-sign.  Each of these 12 yields is a free parameter of
the fit with a penalty term of
\begin{equation}
\label{eq:brpen}
\textrm{Penalty} \;=\;\, 0.5 \;
\frac{\left[Y_\textrm{fit}\left(D^0\!\rightarrow\!\texttt{ref}\right) - 
Y_0\left(D^0\!\rightarrow\!\texttt{ref}\right)\right]^2}
{\left[\sigma\left(Y_0\left(D^0\!\rightarrow\!\texttt{ref}\right)\right)\right]^2}
\end{equation}
added to the log-likelihood for each yield.  $Y_\textrm{fit}$ is the fitted yield, $Y_\textrm{0}$ 
is the expected yield based on the number of right-sign signal events, the relative efficiency between
signal and the reflection, and the branching ratio between signal and the reflection. The uncertainty on the
expected yield, $\sigma(Y_\textrm{0})$,
is based on the PDG~\cite{pdg} branching ratios inflated to account for effects such as Monte Carlo 
misidentification mismatch and efficiency variation due to mismodeled resonant substructure.
The branching ratio and error used in the fit are shown in Table~\ref{tab:br}.  The use of the penalty
terms does not significantly affect the results as the signal shape is very different from these background
shapes.  The benefit is that correlations between backgrounds are significantly reduced resulting in a 
better behaving fit.  The amount of background due to double misidentification of the right-sign signal
mode is fixed based on the fitted right-sign yield and relative efficiency obtained from Monte Carlo.  
This background accounts for only 132 events in the entire data sample, 19 in the signal region.  The
error on this number, determined from the Monte Carlo fidelity in estimating the $D^0\!\rightarrow\!K^-K^+$ contribution,
is less than 12\%, resulting in a negligible contribution to the systematic error.
\begin{table}
\begin{center}
\caption[Branching ratio penalties]{The PDG04 branching ratios and the values used in the fit to constrain the
various background contributions.  Errors are inflated to account for additional uncertainties such as particle identification and
resonance substructure.  For $\pi^-\pi^+\pi^0$ the errors are inflated due to a discounted measurement of $(3.9\pm1.1)\%$.}
\begin{tabular}{lccc}
\textbf{Decay} & \textbf{PDG BR (\%)} & \textbf{Used BR (\%)} \\
$\Gamma(K^-K^+)/\Gamma(K^-\pi^+)$ & $10.23^{\,+\,0.22}_{\,-\,0.27}$ & $10.26\pm0.32$ \\
$\Gamma(\pi^-\pi^+)/\Gamma(K^-\pi^+)$ & $3.62\pm0.10$ & $3.58\pm0.20$ \\
$\Gamma(\pi^-\pi^+\pi^0)/\Gamma_\textrm{total}$ & $1.1\pm0.4\pm0.2$ & $1.1\pm1.5$ \\
$\Gamma(K^-\pi^+\pi^0)/\Gamma(K^-\pi^+)$ & $342\pm22$ & $342\pm55$ \\
$\Gamma(K^-\ell^+\nu)/\Gamma_\textrm{total}$ & $6.76\pm0.28$ & $6.76\pm0.90$ \\
$\Gamma(\overline{K}\,\!^0\pi^+\pi^-)/\Gamma_\textrm{total}$ & $5.97\pm0.35$ & $5.97\pm0.875$ \\
\end{tabular}
\label{tab:br}
\end{center}
\end{table}

During the course of this analysis it was observed that the Monte Carlo failed to reproduce a low
mass tail from the signal $D^0\!\rightarrow\!K^-\pi^+$ decays.  The program 
\textsc{Photos}~\cite{photos} was used to model QED internal bremsstrahlung processes which occur
during the charm decay.  This addition produces a Monte Carlo signal shape which agrees with the data.
Correlations in the Monte Carlo shapes between the three variables ($M(D^0)$, $Q(D^*)$, $\tau(D^0)$)
were examined.  While the lifetime was independent, the mass and energy release had significant 
correlations.  Therefore, the Monte Carlo shapes are represented as a 2-D shape ($M,Q$) times a 1-D 
shape ($\tau$).  The Monte Carlo shapes for individual modes are more than 50 times larger than the
data sample.

\section{Systematic checks and uncertainties}

Mini Monte Carlo tests of the fit were used to verify the accuracy of the reported fit errors and to
check for possible biases in the fit.  1000 independent data samples were generated from the best fit
to our data.  Each of these data samples were fit as real data.  Distributions of fitted values indicate
no significant bias and accurate fit errors.  Confidence levels based on the distribution of $-\log{\mathcal{L}}$
and $\chi^2$ were found to be 63\% and 27\%, respectively.  A direct measurement of the confidence level
from the $\chi^2$ is impossible due to very few entries in most bins.

Fit variants with different binning, different constraints, and different accounting of the random
background were tried.  No significant differences were observed indicating a consistent fit with no
unaccounted-for resolution effects.  Variations of all the selection
criteria were analyzed with no significant differences in the results.  This is to be expected since
the dependence on the Monte Carlo simulation is rather weak.  For the branching ratios
and asymmetries reported, systematic uncertainties due to fit variants and cut variants were obtained 
from the r.m.s of the variations and then added in quadrature.  95\%~CL limits on the asymmetries were
obtained by scaling the 1-$\sigma$ errors.  When the asymmetry was measured directly in a fit, the
scaling was determined by the ratio of errors obtained from \textsc{Minos} for 
$\Delta \log{\mathcal{L}} = 0.50$ (1-$\sigma$) and $\Delta \log{\mathcal{L}} = 1.92$ (95\%~CL).  The 
scaling is calculated separately for positive and 
negative errors.  When the asymmetry was calculated from other parameters the naive scaling of $1.96$ 
between 1-$\sigma$ and 95\%~CL uncertainties was used.

\section{Fit results}

We perform five types of fits to extract branching ratio, mixing, and \textit{CP} violation information.
These fits are labeled A--E.  Fits which constrain ${x'}^2$ to lie in the physical region ($>$ 0) are
``primed.''

Fit A assumes no mixing and no \textit{CP} violation (${x'}^2 = {y'}=0$ in Eq.~\ref{eq:wsrate}).  
The wrong-sign branching ratio is
measured to be $R_\textrm{WS} = (0.429^{\,+\,0.063}_{\,-\,0.061}\pm 0.027)\%$.  Fit B assumes no mixing
but allows for global \textit{CP} violation (${x'}^2 = {y'} = \phi = A_M = 0$ in Eq.~\ref{eq:wsratecp}).  
The wrong-sign branching ratio and \textit{CP} 
asymmetry are found to be $R_\textrm{WS} = (0.429 \pm 0.063 \pm 0.028)\%$ and
$A_D = 0.18 \pm 0.14 \pm 0.041$.

\begin{figure}
\centerline{\includegraphics[width=2.75in,height=3.2in]{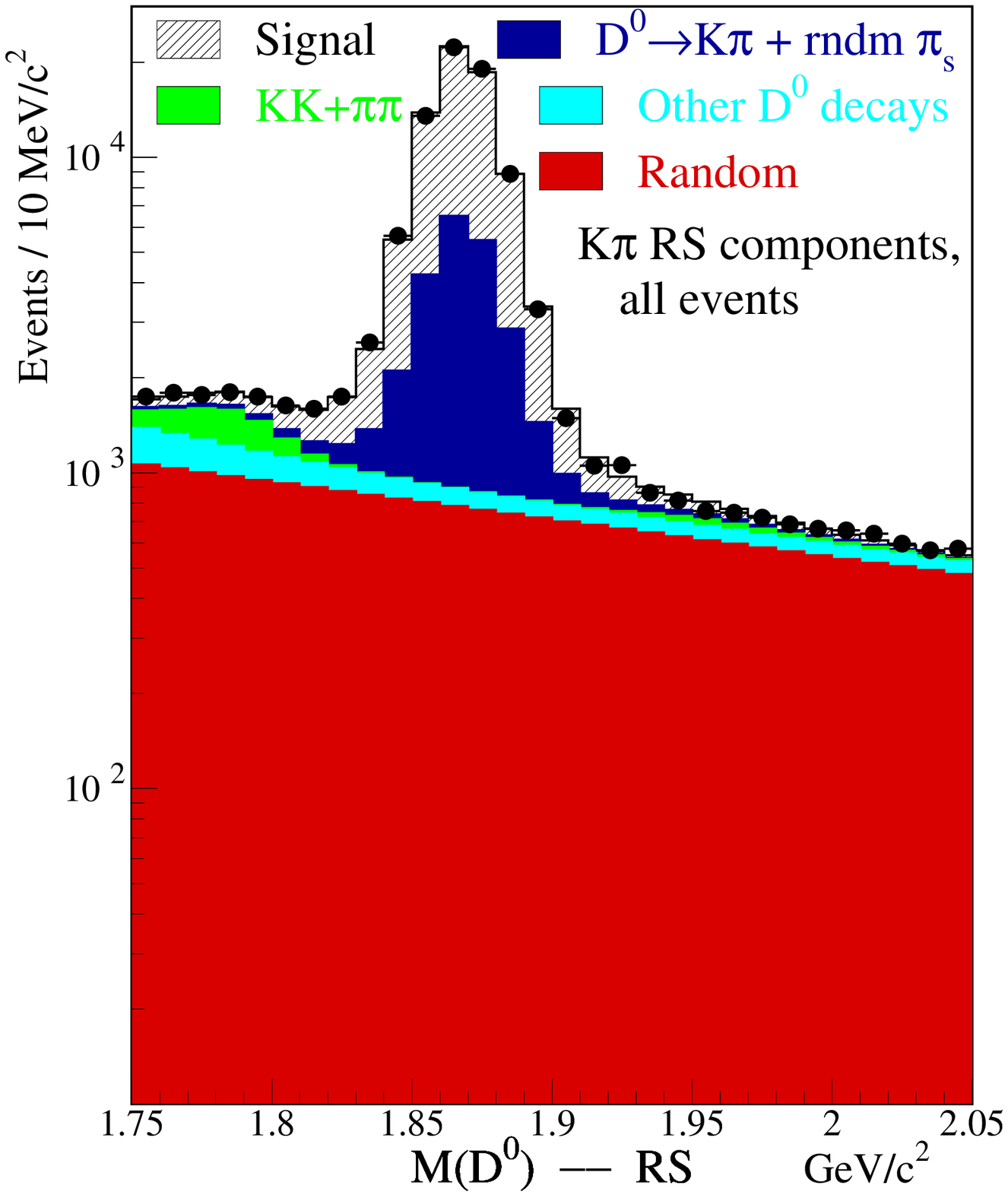}
\includegraphics[width=2.75in,height=3.2in]{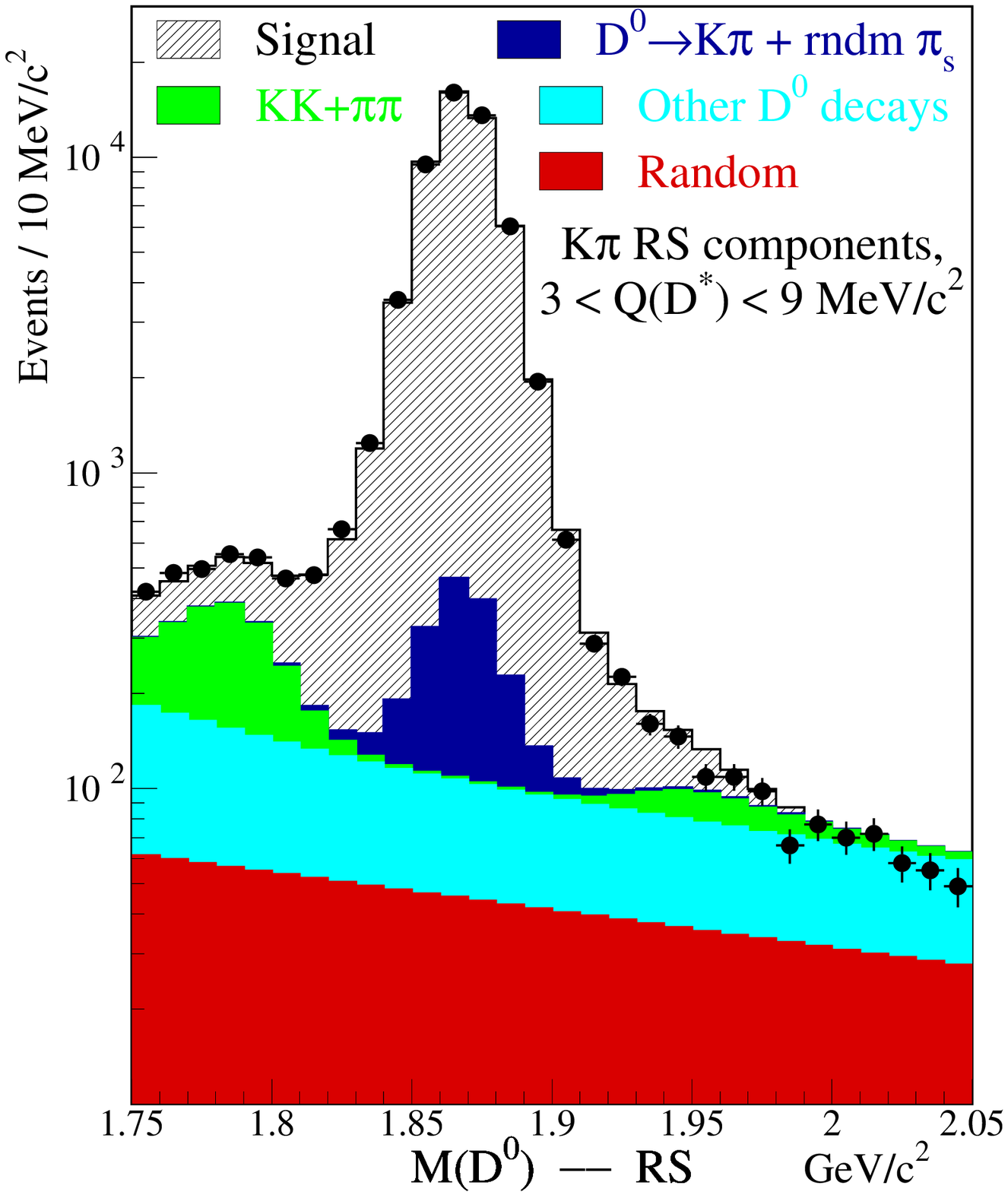}}
\centerline{\includegraphics[width=2.75in,height=3.2in]{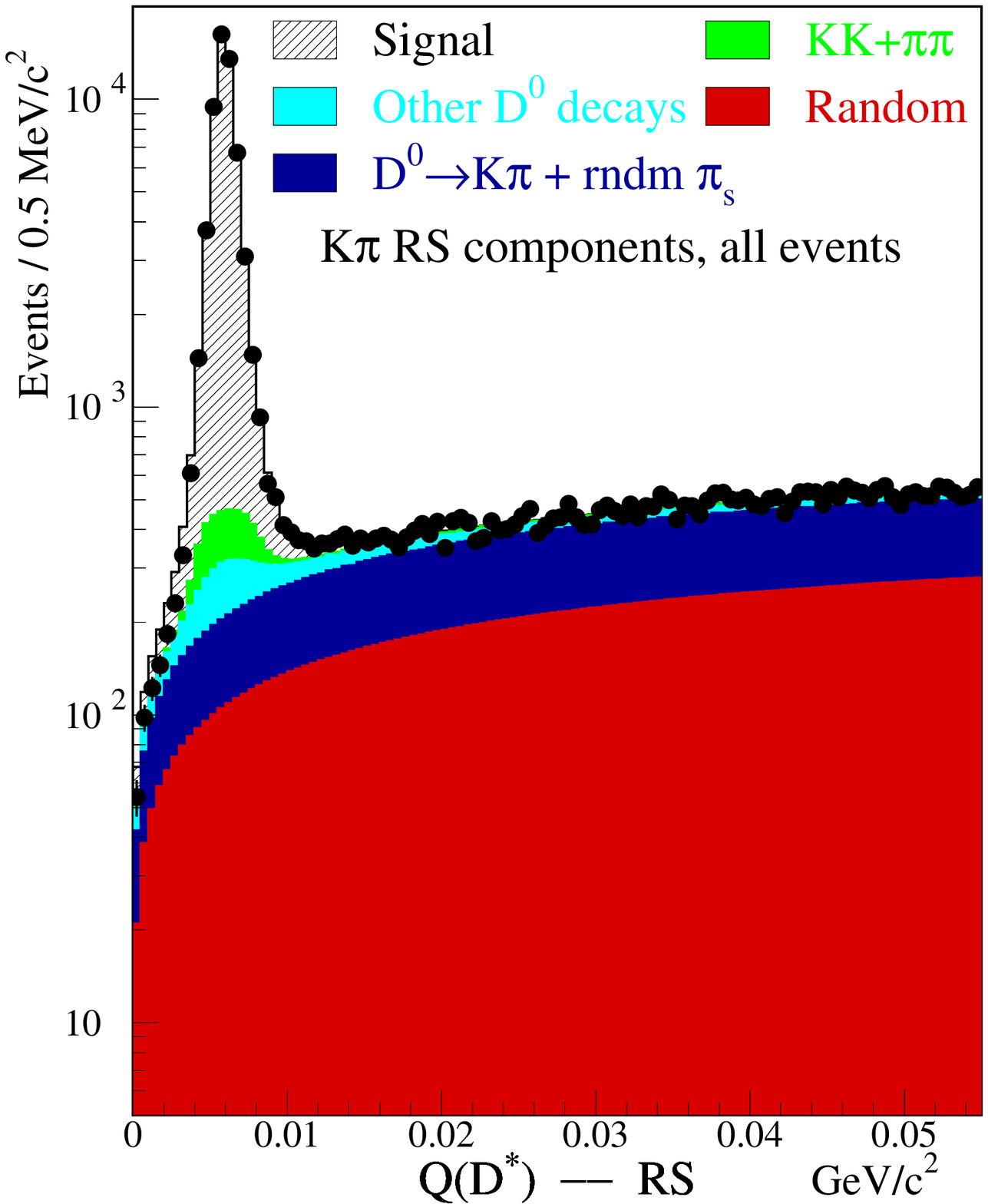}
\includegraphics[width=2.75in,height=3.2in]{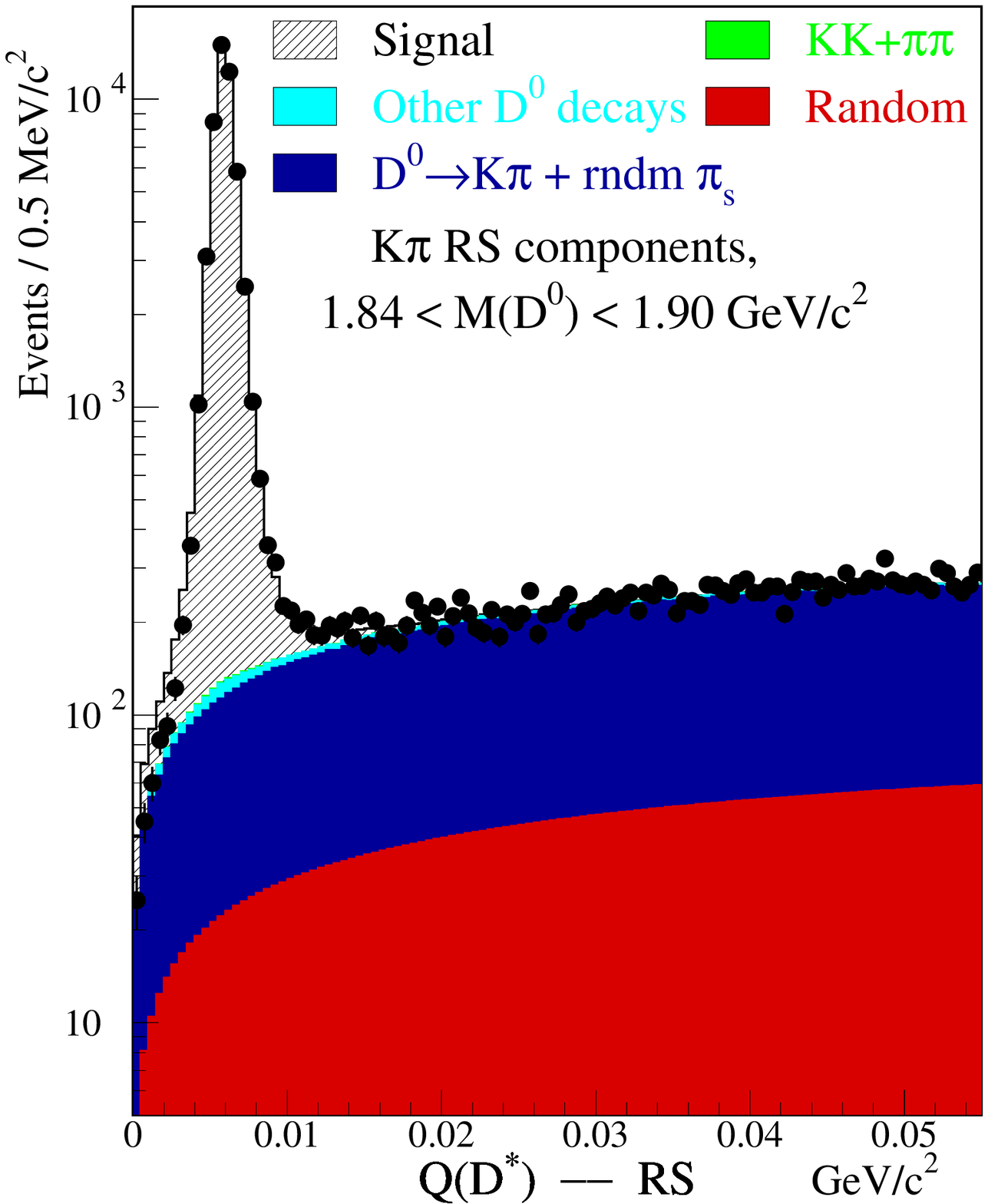}}
\caption{Projection onto $M(D^0)$ and $Q(D^*)$ of both the fit (Fit C)
and data for right-sign
events.  The left plots are for all events while the events in the right plots have 
cuts in the non-plotted variable.  Due to the very large signal-to-background, semi-log
plots are shown.}
\label{fig:mqrs}
\end{figure}

\begin{figure}
\centerline{\includegraphics[width=2.75in,height=3.2in]{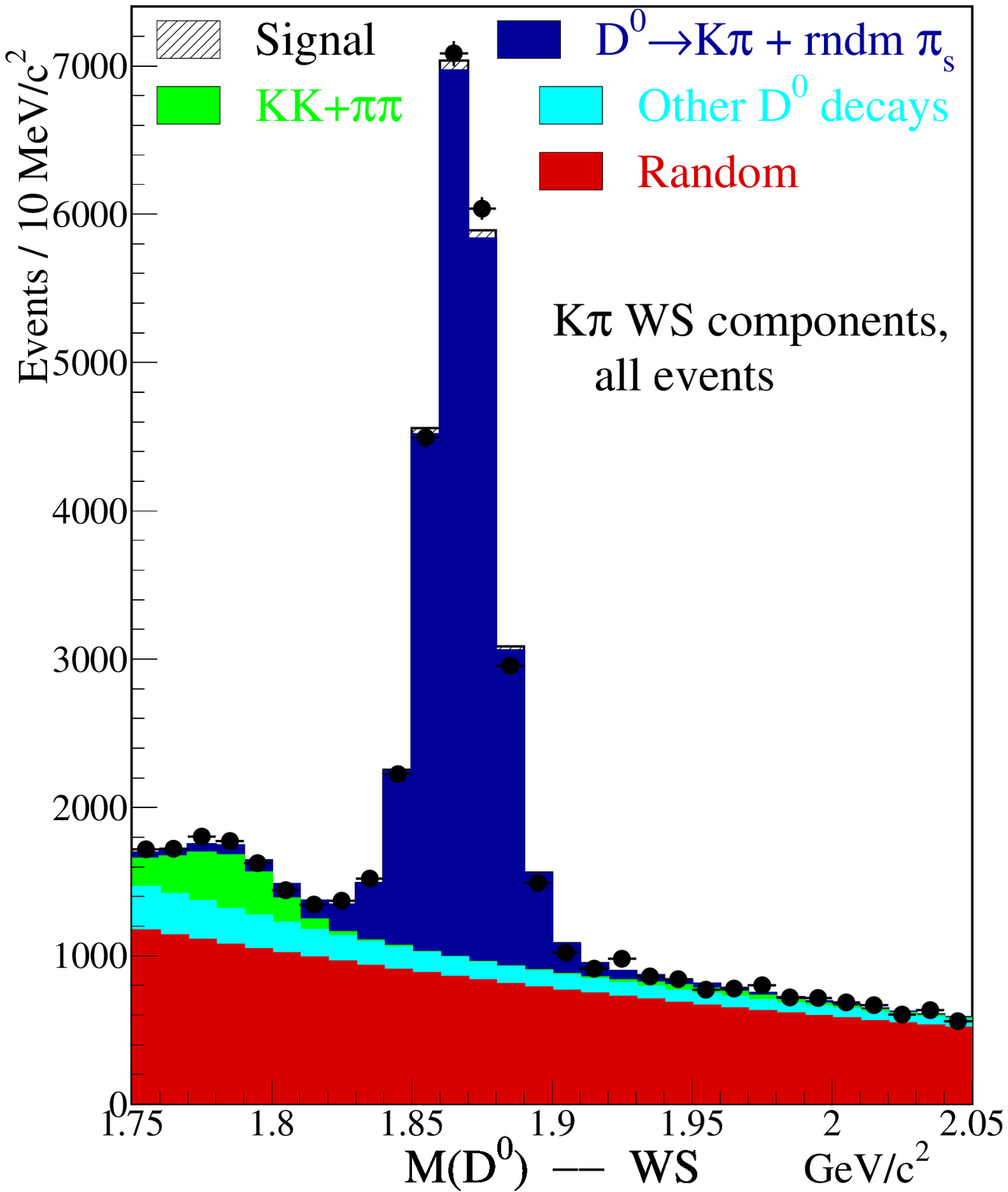}
\includegraphics[width=2.75in,height=3.2in]{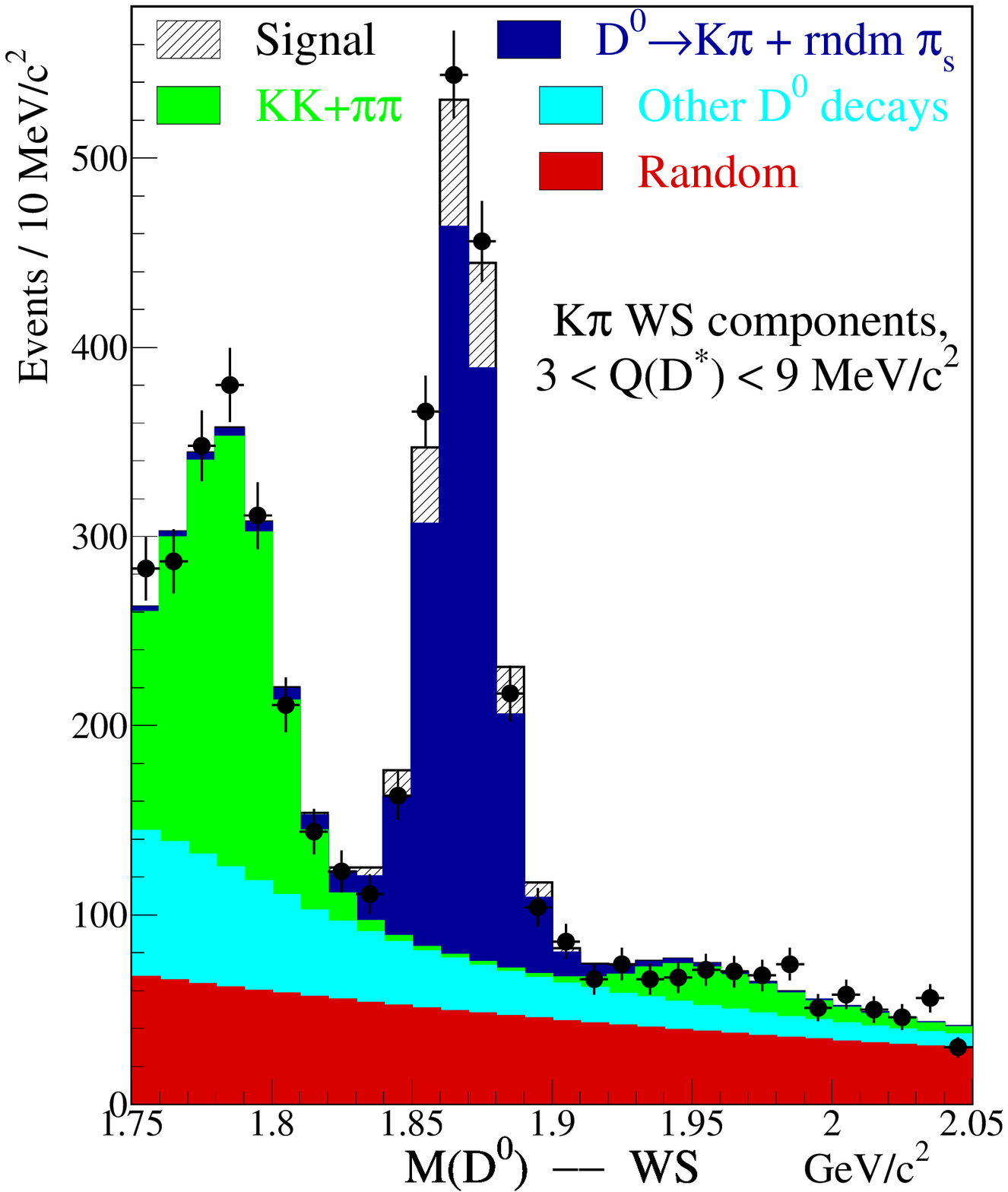}}
\centerline{\includegraphics[width=2.75in,height=3.2in]{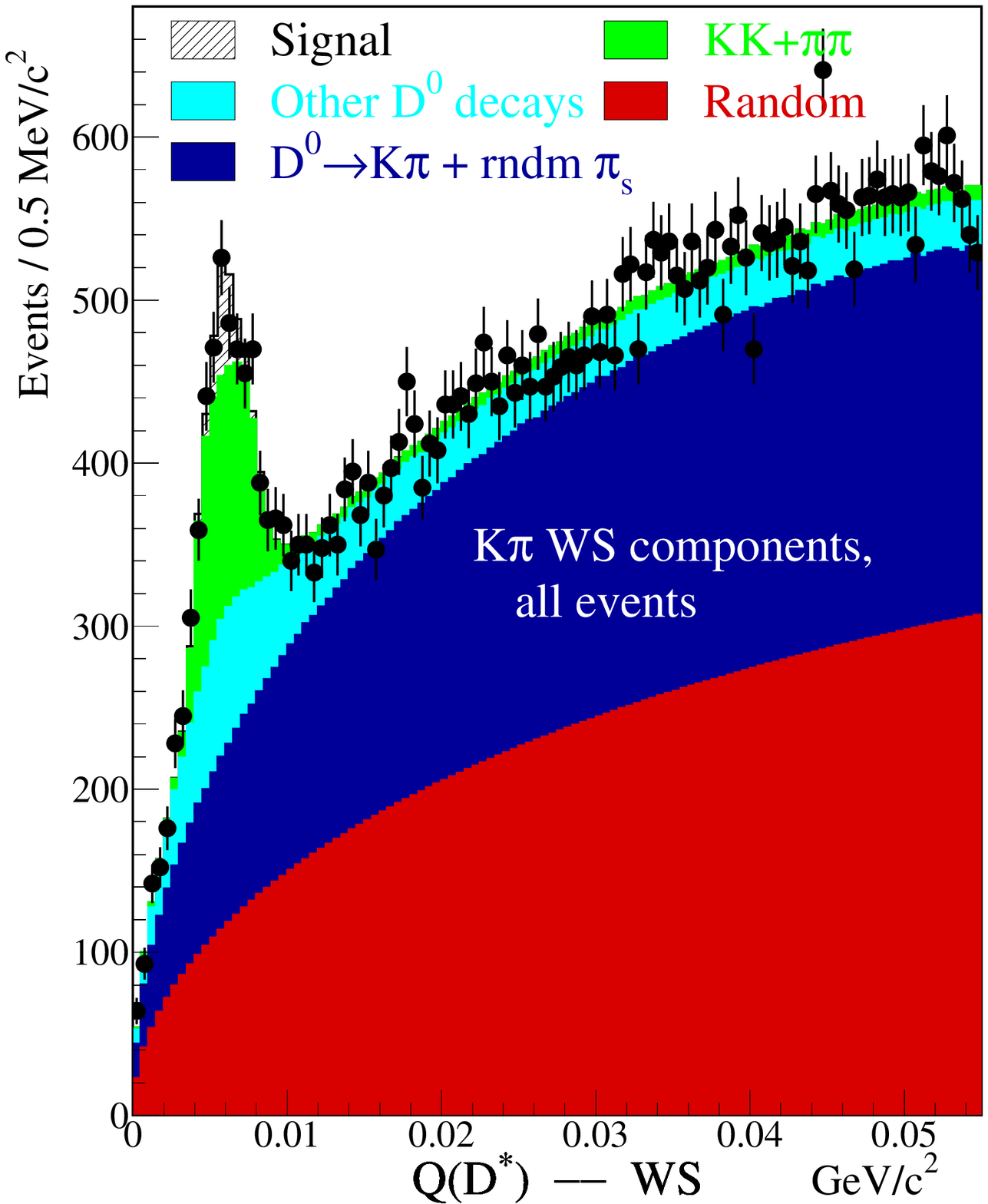}
\includegraphics[width=2.75in,height=3.2in]{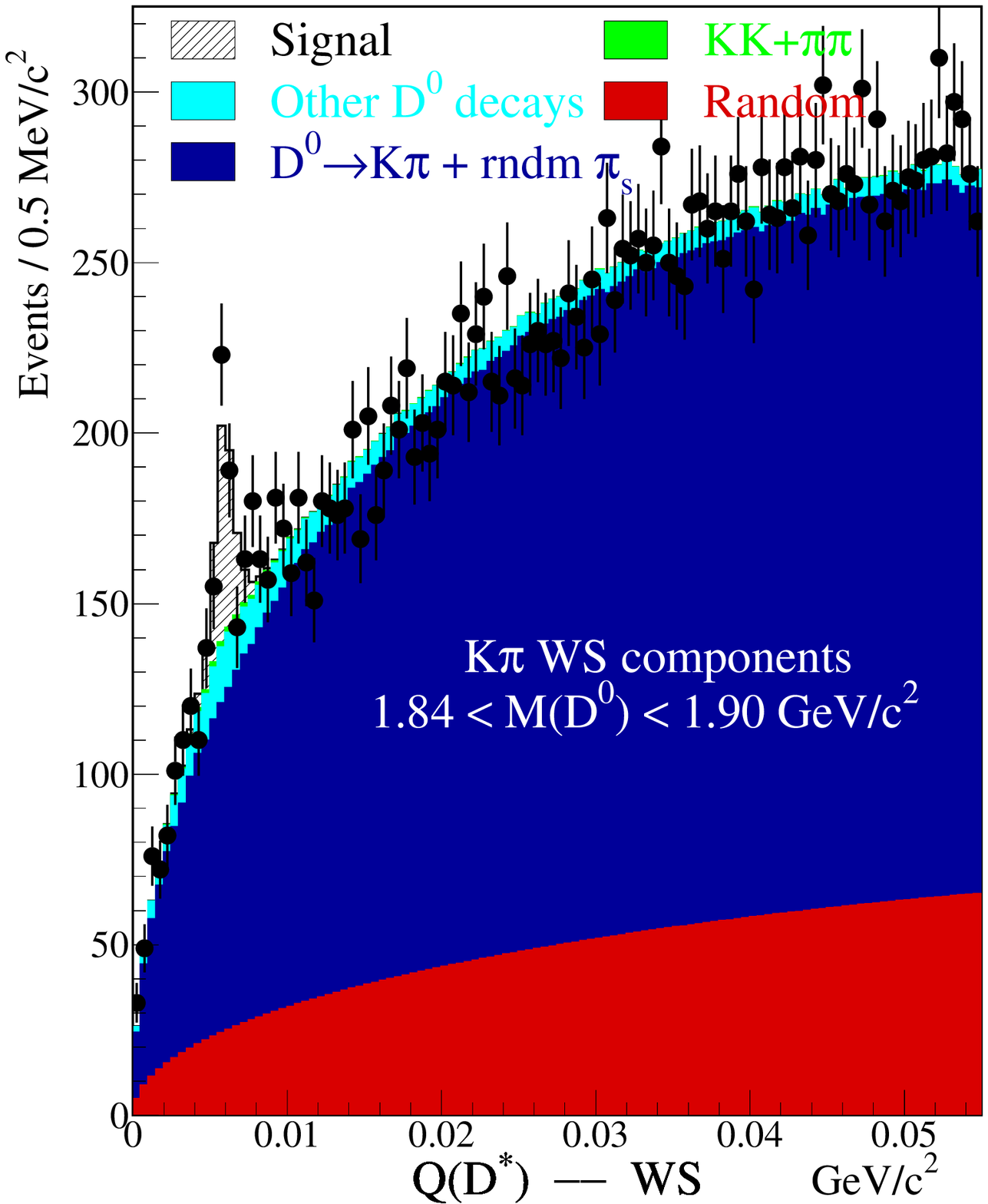}}
\caption{Projection onto $M(D^0)$ and $Q(D^*)$ of both the fit (Fit C) 
and data for wrong-sign
events.  The left plots are for all events while the events in the right plots have 
cuts in the non-plotted variable.}
\label{fig:mqws}
\end{figure}

\begin{figure}
\centerline{\includegraphics[width=2.75in,height=2.8in]{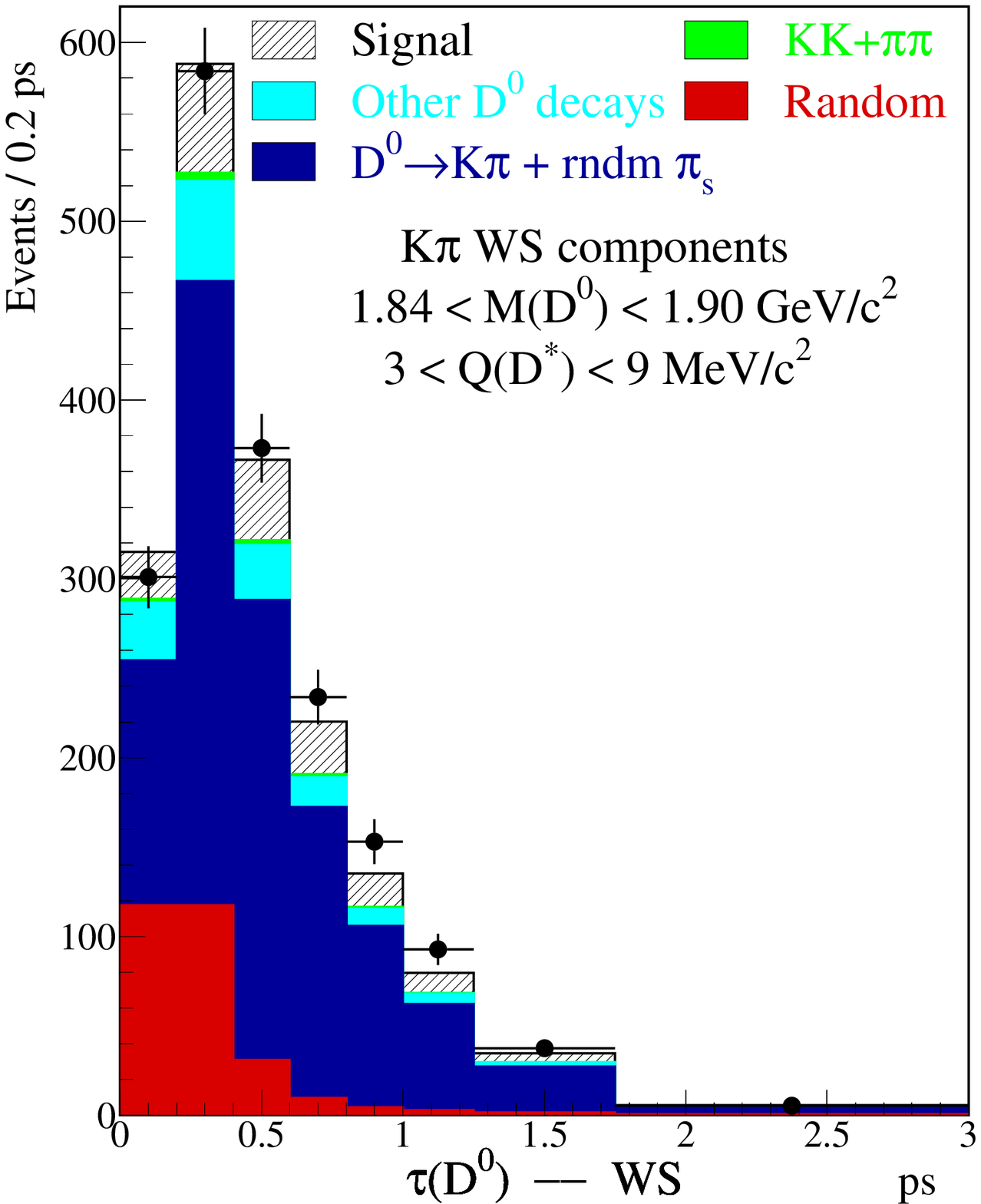}
\includegraphics[width=2.75in,height=2.8in]{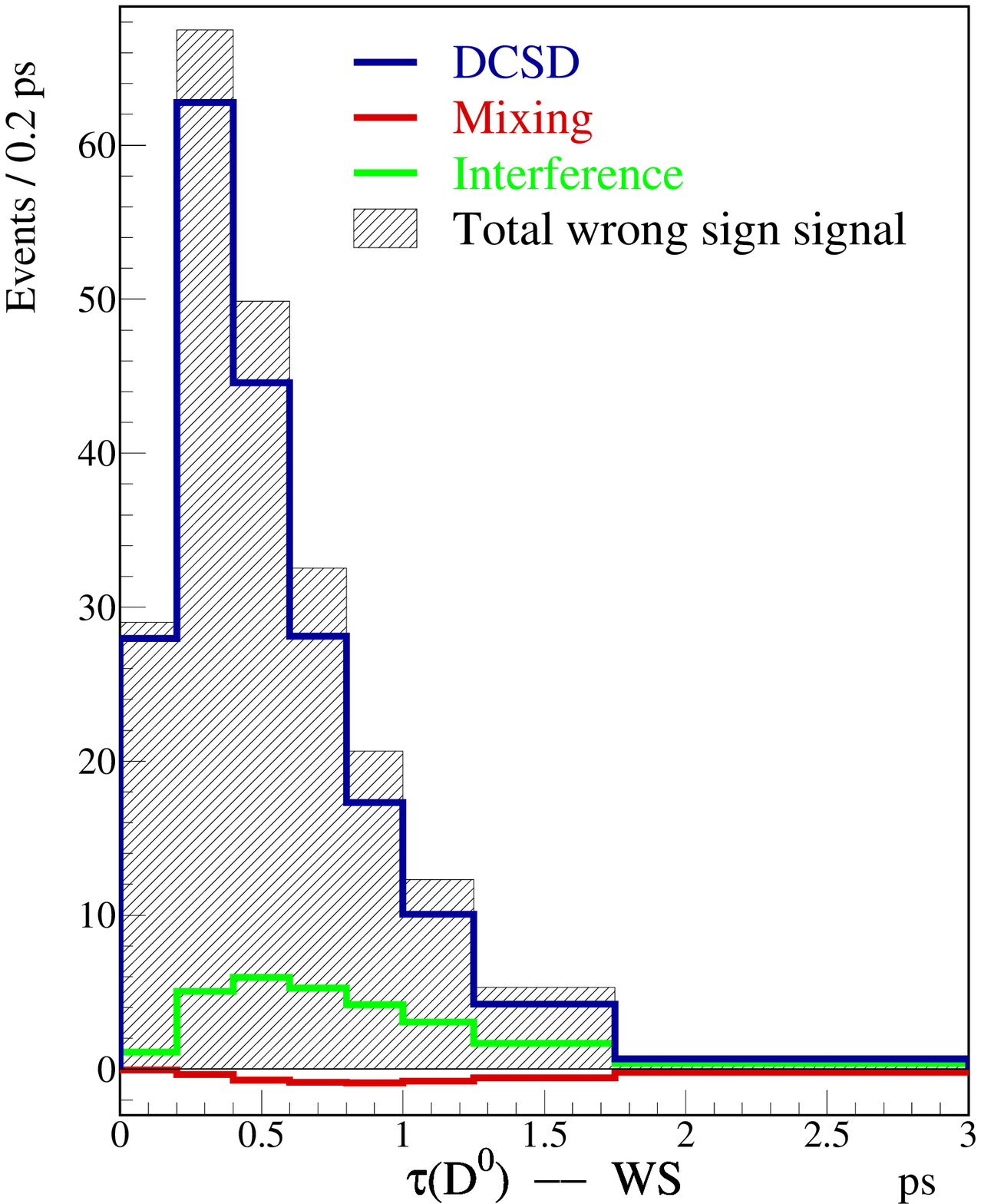}}
\caption{Projection onto $\tau(D^0)$ of the fit (Fit C)
and data for wrong-sign events.
The left plot is for events in the $M(D^0)$ and $Q(D^*)$ signal region while the 
right plot shows the contributions to the wrong-sign signal. The first bin is low 
due to the necessary inefficiency of vertex separation requirements.  The mixing
component is negative due to the unphysical value of ${x'}^2=-0.006$}
\label{fig:t}
\end{figure}

Fits C and C$^\prime$ allow for mixing without \textit{CP} violation (Eq.~\ref{eq:wsrate}).
Representative plots of Fit C are
are shown in Figs.~\ref{fig:mqrs}--\ref{fig:t}.
Figure~\ref{fig:mqrs} show the right-sign projections onto $M(D^0)$ and $Q(D^*)$, with and without 
a cut on $Q(D^*)$ and $M(D^0)$, respectively.  Figure~\ref{fig:mqws} shows the same projections for
the wrong-sign events.
Figure~\ref{fig:t} shows the signal component of the wrong-sign lifetime 
distribution split into the three components, DCSD, mixing, and interference.

\begin{figure}
\centerline{\includegraphics[width=4.0in,height=3.5in]{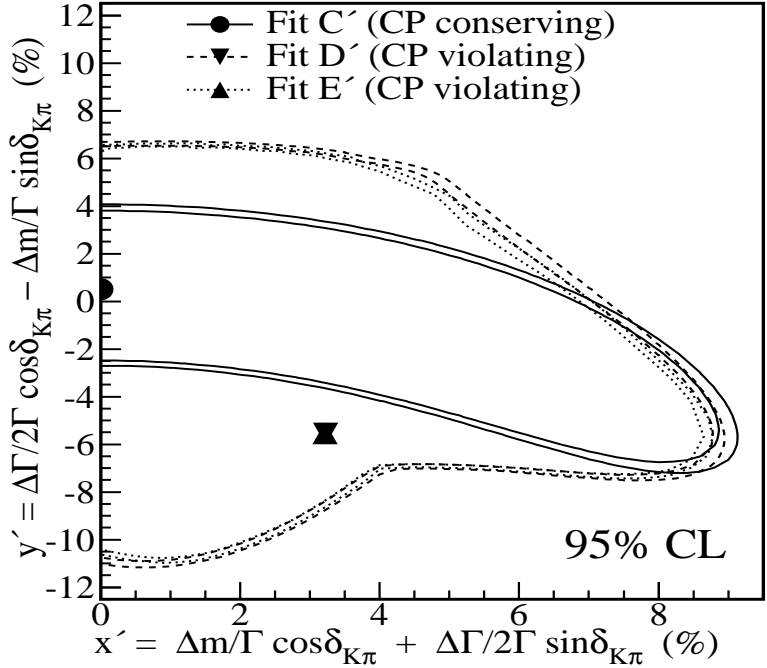}}
\caption{95\%~CL ${x'},{y'}$ contours for \textit{CP} conserving and \textit{CP} violating fits with 
${x'}^2$ constrained to be greater than zero.
The inner curves show the 95\%~CL contours including statistical uncertainties only while the 
outer curves also include systematic uncertainties.}
\label{fig:contour}
\end{figure}

The DCS branching ratio is found
to be $R_D = (0.381^{\,+\,0.167}_{\,-\,0.163}\pm0.092)\%$.  The negative log-likelihood is minimized
at values of ${x'}^2 = -0.059\%$ and $y' = 1.0\%$ for Fit C\@.  Due to significant correlations between 
${x'}$ and $y'$ and the fact that ${x'}^2 < 0$ is unphysical, the interesting physics result is
a contour in the ${x'}$, $y'$ plane.  The minimum negative 
log-likelihood with ${x'}^2 \ge 0$ occurs at ${x'}^2 = 0$ and $y' = 0.5\%$. 
Although we fit for ${x'}^2$, we choose to plot ${x'}$.
The 95\%~CL contour is defined as the location in the 
$x'$, $y'$ plane where the change in log-likelihood reaches $3.00$ relative to the minimum 
negative log-likelihood in the physical region of the $x'$, $y'$ plane.  This method has been checked
with a frequentist method using mini Monte Carlo with identical results.
Systematic checks were
performed with 120 fit and cut variants.  The contour variation is consistent with differences in the
returned value of ${x'}^2$ and $y'$.  To assess the systematic uncertainty, we first determine the change 
in negative log-likelihood between the global minimum and the ${x'}$, $y'$ location for each of the
120 variants.  We then find the value greater than 95\% of these differences which is $0.48$.  We
find the contour at which the change in negative log-likelihood is $3.00+0.48 = 3.48$ and
call this the 95\%~CL including systematic uncertainties.  The contours for Fit C$^\prime$ are shown in 
Fig.~\ref{fig:contour}.  We define 95\%~CL limits on $x'$ and $y'$ based on the projection of the
contour onto the respective axis.  The 95\%~CL limit on $R_M$ is defined as the maximum $R_M$ value
on the contour. The improvement in $-\log{\mathcal{L}}$ compared to Fit A is 0.089 (0.082) for
free (restricted) values of ${x'}^2$.

The last two fits allow for both mixing and \textit{CP} violation.  We do not have sufficient statistics
to obtain useful measurements of the mixing and interference \textit{CP} violating parameters $A_M$ and $\sin{\phi}$
(although they are left free in the fit).
Thus, the primary results are the DCSD \textit{CP} violating parameter $A_D$ and the mixing parameters. 
The Fit D and D$^\prime$ results come from a fit to the combined
$D^0$ and $\overline{D}\,\!^0$ sample using Eq.~\ref{eq:wsratecp}.  The global minimum for Fit D 
occurs at
a highly unphysical ${x'}^2 = -0.52\%$ and $y' = 6.6\%$ which makes the nominal branching ratio 
uninteresting.  The minimum in the physical plane (Fit D$^\prime$) occurs at ${x'}^2 = 0.023\%$ and $y' = -2.6\%$
with an increase in the log-likelihood fo $0.94$ compared to Fit D\@.
The branching ratio and \textit{CP} asymmetry for the physical fit are found to be 
$R_D = (0.521^{\,+\,0.144}_{\,-\,0.138} \pm 0.076)\%$ and 
$A_D = 0.13^{\,+\,0.33}_{\,-\,0.26} \pm 0.10$.
The statistical and statistical+systematic contours are determined in the same manner as in Fit C, 
as are the 95\%~CL limits on the mixing parameters.  A mini Monte Carlo study using the best fit function
to create fake data sets was employed to check the 95\%~CL contour.  Only $92.8\pm0.8\%$ of the fake data
sets had fitted $x',y'$ values inside the contour.  Therefore the contour was increased to include 95\% of
the fake data sets by increasing $\Delta \log{\mathcal{L}}$ from $3.00$ to $3.50$.
The Fit D$^\prime$ contours are shown in Fig.~\ref{fig:contour}.  The improvement in $-\log{\mathcal{L}}$ of
Fit D (D$^\prime$) compared to Fit B is 2.51 (1.57).

The last fits, E and E$^\prime$, fit the $D^0$ and $\overline{D}\,\!^0$ samples separately using 
Eq.~\ref{eq:wsrate}.  From these fits we can calculate $R_D = \sqrt{R_D^+R_D^-} = (0.518^{\,+\,0.152}_{\,-\,0.145} \pm 0.076)\%$ and
$A_D = 0.12^{\,+\,0.29}_{\,-\,0.28} \pm 0.10$.
Contours at the $1- \sqrt{0.05} = 77.6\%$ ($\Delta \log{\mathcal{L}} = 1.50$) are 
obtained for each flavor.
Every point in the $D^0$ contour is combined with every point in the $\overline{D}\,\!^0$ 
contour.  The 95\%~CL $x'$, $y'$ contour is defined by the outer edge of the resulting points.  
The systematic error contour is constructed
in a fashion similar to Fits C$^\prime$ and D$^\prime$\@.  
The individual $D^0$ and $\overline{D}\,\!^0$ contours
are inflated to $\Delta \log{\mathcal{L}} = 1.50 + 0.12 = 1.62$ and the combined contour found as before.  
A mini Monte Carlo study using fake data sets created from the best fit to the data was performed to check the
$77.6\%$~CL\@.  The $D^0$ $(\overline{D}{}^0)$ contour was found to encompass significantly less (more) than $77.6\%$ of
the fake data sets.  Therefore, the contours were each adjusted to include $77.6\%$ of the fake data sets by changing
$\Delta \log{\mathcal{L}}$ from the default $1.50$.  After this adjustment, the constructed 95\%~CL limit was found
to be consistent with including 95\% of the fake data sets.
The Fit E$^\prime$ contours are shown in Fig.~\ref{fig:contour}. 
Limits on the mixing parameters are set as in Fits C$^\prime$ and D$^\prime$ by projecting the contour onto the
axes.

\section{Conclusions}

A compendium of the results from this analysis can be found in 
Tables~\ref{tab:results_abc} and \ref{tab:results_de}.  
The contours obtained in this analysis~\cite{focus_contour}
agree better with recent results from BABAR~\cite{babar_hmix} and BELLE~\cite{belle_hmix} 
than with older results from CLEO~\cite{cleo_hmix}.
The no mixing, no \textit{CP} violation result agrees with the previous FOCUS result 
obtained from the same parent data sample but utilizing a very different technique~\cite{linkpaper}.
The two different techniques used to obtain the mixing with \textit{CP} violation results, 
Fits $D^{(\prime)}$ and $E^{(\prime)}$, return essentially the same answer. 
Fit $D^{(\prime)}$ is to be preferred, however, since it retains all correlations in the fit.
The results reported here represent the best published charm mixing limit from 
a fixed-target experiment.  In some respects, this measurement is complementary to that obtained from
the cleaner, higher statistics $e^+e^-$ collider experiments.  FOCUS has the world's most 
accurate lifetime measurements for the six charm species it has measured.  
In addition, the FOCUS lifetime resolution is far superior to that of the $e^+e^-$ collider experiments.  
These strengths, along with significant differences in production and reconstruction, result in a very
different analysis from those of the $e^+e^-$ collider experiments, providing a valuable check on 
those results.

\begin{table}
\begin{center}
\caption{Summary of results for Fits A--D, described in the text.}
\label{tab:results_abc}
\begin{tabular}{lcc}
Measurement & Value & 95\% CL limit \\ \hline \hline
\multicolumn{3}{c}{Fit A: No mixing, \textit{CP} conserving} \\
$R_\textrm{WS}$ & $(0.429^{\,+\,0.063}_{\,-\,0.061}\pm 0.027)\%$ \\ \hline
\multicolumn{3}{c}{Fit B: No mixing, \textit{CP} violating} \\
$R_\textrm{WS}$ & $(0.429 \pm 0.063 \pm 0.028)\%$ \\
$A_D$ & $0.18 \pm 0.14 \pm 0.04$ & $-0.11 < A_D < 0.48$ \\ \hline
\multicolumn{3}{c}{Fit C: Mixing, \textit{CP} conserving, no ${x'}^2$ constraint} \\
$R_D$ & $(0.381^{\,+\,0.167}_{\,-\,0.163}\pm0.092)\%$ \\
${x'}^2$ & $-0.059\%$ \\
$y'$ & $1.0\%$ \\ \hline
\multicolumn{3}{c}{Fit C$^\prime$: Mixing, \textit{CP} conserving, ${x'}^2 > 0$ constraint} \\
$R_D$ & $(0.395^{\,+\,0.154}_{\,-\,0.098}\pm 0.069)\%$ \\
$R_M$ & & $< 0.63\%$ \\
${x'}^2$ & $0.00\%$ & $<0.83\%$ \\
$y'$ & $0.5\%$ & $-7.2\% < y' < 4.1\%$ \\ \hline
\multicolumn{3}{c}{Fit D: Mixing, \textit{CP} violating, combined $D^0$ and $\overline{D}{}^0$ fit, no ${x'}^2$ constraint} \\
$R_D$ & $(0.255^{\,+\,0.126}_{\,-\,0.145} \pm 0.132)\%$ \\
$A_D$ & $0.06^{\,+\,0.46}_{\,-\,0.65} \pm 0.13$ \\
${x'}^2$ & $-0.52\%$ \\
$y'$ & $6.6\%$ \\ \hline
\multicolumn{3}{c}{Fit D$^\prime$: Mixing, \textit{CP} violating, combined $D^0$ and $\overline{D}{}^0$ fit, ${x'}^2 > 0$ constraint} \\
$R_D$ & $(0.517^{\,+\,0.147}_{\,-\,0.158} \pm 0.076)\%$ \\
$A_D$ & $0.13^{\,+\,0.33}_{\,-\,0.25} \pm 0.10$ & $-0.42 < A_D < 0.81$ \\
$R_M$ & & $< 0.63\%$ \\
${x'}^2$ & $0.023\% $ & $<0.80\%$ \\
$y'$ & $-2.6\% $ & $-11.2\% < y' < 6.7\%$ \\ \hline \hline
\end{tabular}
\end{center}
\end{table}

\begin{table}
\begin{center}
\caption{Summary of results for Fit E, described in the text.}
\label{tab:results_de}
\begin{tabular}{lcc}
Measurement & $D^0$ Result & $\overline{D}{}^0$ Result \\ \hline \hline
\multicolumn{3}{c}{Fit E: Mixing, \textit{CP} violating, separate $D^0$ and $\overline{D}{}^0$ fits, no ${x'}^2$ constraint} \\
$R_D^\pm$ & $(0.275^{\,+\,0.231}_{\,-\,0.233} \pm 0.117)\%$ & $(0.458^{\,+\,0.235}_{\,-\,0.226} \pm 0.099)\%$ \\
$R_D$ & \multicolumn{2}{c}{$(0.355^{\,+\,0.175}_{\,-\,0.174} \pm 0.107)\%$} \\
$A_D$ & \multicolumn{2}{c}{$-0.25 \pm 0.46 \pm 0.14$} \\
${x'}^{\pm^2}$ & $-2.12\%$ & $0.61\%$ \\
${y'}^\pm$ & $9.88\%$ & $-4.43\%$ \\ \hline
\multicolumn{3}{c}{Fit E$^\prime$: Mixing, \textit{CP} violating, separate $D^0$ and $\overline{D}{}^0$ fits, ${x'}^2 > 0$ constraint} \\
$R_D^\pm$ & $(0.586^{\,+\,0.169}_{\,-\,0.157} \pm 0.079)\%$ & $(0.458^{\,+\,0.235}_{\,-\,0.226} \pm 0.099)\%$ \\
$R_D$ & \multicolumn{2}{c}{$(0.518^{\,+\,0.152}_{\,-\,0.145} \pm 0.076)\%$} \\
$A_D$ & \multicolumn{2}{c}{$0.12^{\,+\,0.29}_{\,-\,0.28} \pm 0.10$} \\
$A_D$ & \multicolumn{2}{c}{$-0.46 < A_D < 0.72$ @ 95\% CL} \\
$R_M$ & \multicolumn{2}{c}{$< 0.61\%$ @ 95\% CL} \\
${x'}^{\pm^2}$ & $0.00\%$ & $0.61\%$ \\
${y'}^\pm$ & $-1.02\%$ & $-4.43\%$ \\
${x'}^2$ & \multicolumn{2}{c}{$0.023\%$,~~~~ $<0.77\%$ @ 95\% CL} \\
$y'$ & \multicolumn{2}{c}{$-2.6\%$,~~~~ $-11.0\% < y' < 6.6\%$ @ 95\% CL} \\ \hline \hline
\end{tabular}
\end{center}
\end{table}

\section{Acknowledgments}
We wish to acknowledge the assistance of the staffs of Fermi National
Accelerator Laboratory, the INFN of Italy, and the physics departments
of the collaborating institutions. This research was supported in part
by the U.~S.  National Science Foundation, the U.~S. Department of
Energy, the Italian Istituto Nazionale di Fisica Nucleare and
Ministero dell'Istruzione dell'Universit\`a e della Ricerca, the
Brazilian Conselho Nacional de Desenvolvimento Cient\'{\i}fico e
Tecnol\'ogico, CONACyT-M\'exico, the Korean Ministry of Education, 
and the Korean Science and Engineering Foundation.

\bibliographystyle{unsrt}

\end{document}

%% file: authors.tex
\collaboration{The FOCUS Collaboration}

\author[ucd]{J.~M.~Link}
\author[ucd]{P.~M.~Yager}
\author[cbpf]{J.~C.~Anjos}
\author[cbpf]{I.~Bediaga}
\author[cbpf]{C.~G\"obel}
\author[cbpf]{A.~A.~Machado}
\author[cbpf]{J.~Magnin}
\author[cbpf]{A.~Massafferri}
\author[cbpf]{J.~M.~de~Miranda}
\author[cbpf]{I.~M.~Pepe}
% ...

\address[ucd]{University of California, Davis, CA 95616}
\address[cbpf]{Centro Brasileiro de Pesquisas F\'{\i}sicas, Rio de Janeiro, RJ, Brasil}
% ...
